\documentclass[twocolumn]{aastex631}
\usepackage[utf8]{inputenc}

\usepackage{graphicx}
\usepackage{xfrac}
\usepackage{mathrsfs}
\usepackage{soul}

\label{key}

\newcommand{\spt}{SPT2349$-$56}

\received{XXX}
\revised{YYY}
\accepted{ZZZ}
\submitjournal{ApJ}

\shorttitle{Radio-AGN in SPT2349}
\shortauthors{Chapman et al.}
\graphicspath{{./}{figures/}}

\begin{document}

\title{ 
An Overabundance of Radio-AGN in the SPT2349$-$56 Protocluster: Preheating the Intra-Cluster Medium}

\correspondingauthor{Scott C. Chapman}
\email{scott.chapman@dal.ca}

\author[0000-0002-8487-3153]{Scott C.\ Chapman}
\affiliation{Department of Physics and Atmospheric Science, Dalhousie University, Halifax, NS, B3H 4R2, Canada}
\affiliation{NRC Herzberg Astronomy and Astrophysics, 5071 West Saanich Rd, Victoria, BC, V9E 2E7, Canada}
\affiliation{Department of Physics and Astronomy,  University of British Columbia, Vancouver, BC, V6T1Z1, Canada}
\affiliation{Eureka Scientific Inc, Oakland, CA 94602, USA}
\author[0000-0003-1027-5043]{Roger P.\ Deane}
\affiliation{Wits Centre for Astrophysics, School of Physics, University of the Witwatersrand, 1 Jan Smuts Avenue, Johannesburg, 2000, South Africa}
\affiliation{Department of Physics, University of Pretoria, Private Bag X20, Pretoria 0028, South Africa}
\author{Dazhi Zhou}
\affiliation{Department of Physics and Astronomy,  University of British Columbia, Vancouver, BC, V6T1Z1, Canada}
\author[0000-0002-6290-3198]{Manuel Aravena}
\affiliation{N\'ucleo de Astronomia, Facultad de Ingenieria y Ciencias, Universidad Diego Portales, Av. Ej\'ercito 441, Santiago, Chile}
\author{William Rasakanya}
\affiliation{Wits Centre for Astrophysics, School of Physics, University of the Witwatersrand, 1 Jan Smuts Avenue, Johannesburg, 2000, South Africa}
\author[0000-0002-0517-9842]{Melanie Archipley}
\affiliation{Department of Astronomy and Astrophysics, University of Chicago,
5640 South Ellis Avenue, Chicago, IL, 60637, USA}
\affiliation{Kavli Institute for Cosmological Physics, University of Chicago, 5640
South Ellis Avenue, Chicago, IL, 60637, USA}

\author{James Burgoyne}
\affiliation{Department of Physics and Astronomy, University of Victoria, Victoria, Canada.}

\author{Jared Cathey}
\affiliation{Department of Astronomy, University of Florida, 211 Bryant Space Science Center, Gainesville, FL 32611-2055, USA}

\author[0000-0002-0933-8601]{Anthony H.\ Gonzalez}
\affiliation{Department of Astronomy, University of Florida, 211 Bryant Space Science Center, Gainesville, FL 32611-2055, USA}

\author{Ryley Hill}
\affiliation{Department of Physics and Astronomy,  University of British Columbia, Vancouver, BC, V6T1Z1, Canada}

\author{Chayce Hughes}
\affiliation{Department of Physics and Astronomy,  University of British Columbia, Vancouver, BC, V6T1Z1, Canada}

\author{M\`{o}nica Natalia Isla Llave}
\affiliation{INAF Osservatorio di Astrofisica e Scienza dello Spazio di Bologna (OAS), Via Gobetti 93/3, I-40129 Bologna, Italy}
\affiliation{Dipartimento di Fisica e Astronomia (DIFA), Universit\`{a} di Bologna, via Gobetti 93/2, I-40129 Bologna, Italy}

\author{Matt Malkan}
\affiliation{University of California, Los Angeles, Department of Physics and Astronomy, 430 Portola Plaza, Los Angeles, CA 90095, USA}

\author[0000-0001-7946-557X]{Kedar~A. Phadke}
\affil{Department of Astronomy, University of Illinois, 1002 West Green St., Urbana, IL 61801, USA}
\affil{Center for AstroPhysical Surveys, National Center for Supercomputing Applications, 1205 West Clark Street, Urbana, IL 61801, USA}
\affil{NSF-Simons AI Institute for the Sky (SkAI), 172 E. Chestnut St., Chicago, IL 60611, USA}

\author{Vismaya Pillai}
\affiliation{Department of Physics and Astronomy,  University of British Columbia, Vancouver, BC, V6T1Z1, Canada}

\author{Ana Posses}
\affiliation{Department of Physics and Astronomy and George P. and Cynthia Woods Mitchell Institute for Fundamental Physics and Astronomy, Texas A\&M University, 4242 TAMU, College Station, TX 77843-4242, US}

\author{Bonnie Slocombe}
\affiliation{Department of Physics and Astronomy,  University of British Columbia, Vancouver, BC, V6T1Z1, Canada}

\author{Manuel Solimano}
\affiliation{Centro de Astrobiolog\'ia (CAB), CSIC-INTA, Ctra.\ de Ajalvir km 4, Torrej\'on de Ardoz, E-28850, Madrid, Spain}

\author[0000-0003-3256-5615]{Justin Spilker}
\affiliation{Department of Physics and Astronomy and George P. and Cynthia Woods Mitchell Institute for Fundamental Physics and Astronomy, Texas A\&M University, 4242 TAMU, College Station, TX 77843-4242, US}

\author[0000-0002-9181-9948]{Nikolaus Sulzenauer}
\affiliation{Max-Planck-Institut f\"{u}r Radioastronomie, Auf dem Hugel 69, Bonn, D-53121, Germany}

\author{Fabio Vito}
\affiliation{INAF Osservatorio di Astrofisica e Scienza dello Spazio di Bologna (OAS), Via Gobetti 93/3, I-40129 Bologna, Italy}

\author[0000-0001-7192-3871]{Joaquin D. Vieira}
\affiliation{Department of Astronomy, University of Illinois, 1002 West Green St., Urbana, IL 61801, USA}
\affiliation{Center for AstroPhysical Surveys, National Center for Supercomputing Applications, 1205 West Clark Street,
Urbana, IL 61801, USA}
\author{David Vizgan}
\affiliation{Department of Astronomy, University of Illinois, 1002 West Green St., Urbana, IL 61801, USA}

\author{George Wang}
\affiliation{Department of Physics and Astronomy,  University of British Columbia, Vancouver, BC, V6T1Z1, Canada}

\author[0000-0003-4678-3939]{Axel Weiss}
\affiliation{Max-Planck-Institut f\"{u}r Radioastronomie, Auf dem Hugel 69, Bonn, D-53121, Germany}


\begin{abstract}
Following the detection of a radio-loud Active Galactic Nucleus (AGN) in the $z$=$4.3$ protocluster \spt, we have obtained additional observations with MeerKAT in S-band (2.4\,GHz)  with the aim of further characterizing radio 
emission from amongst the ${\sim}\,$30 submillimeter (submm) galaxies (SMGs) identified in the structure. 
We newly identify three of the protocluster SMGs individually at 2.4\,GHz as having a radio-excess, 
two of which are now known to be X-ray luminous AGN. Two additional members are also detected  with radio emission consistent with their star formation rate (SFR). 
Archival MeerKAT UHF (816\,MHz) observations further constrain luminosities and radio spectral indices of these five galaxies.
The Australia Telescope Compact Array (ATCA) is used to detect and resolve the central two sources at 5.5 and 9.0\,GHz finding elongated, jet-like morphologies.
The excess radio luminosities range from  $L_{1.4, rest}\,{=}\,(1-20)\,{\times}\,10^{25}$\,W\,Hz$^{-1}$,  $\sim$10--100$\times$ higher than expected from the SFRs,  assuming the usual far-infrared-radio correlation. 
Of the known cluster members, only the SMG `N1'  shows signs of AGN in any other diagnostics, 
namely a large and compact excess in $^{12}$CO(11--10) line emission.
We compare these results to field samples of radio sources and SMGs.  
The overdensity of radio-loud AGN in the compact core region of the cluster may be providing significant heating to the recently discovered nascent intra-cluster medium (ICM) in \spt. 
\end{abstract}



\section{Introduction}
\label{sec:intro}

Protoclusters of galaxies, the precursors of established clusters, exhibit a reversal of the morphology-density relation seen in clusters at $z<1$ \citep[e.g.,][]{dressler1980,postman05}, whereby star forming galaxies are found in abundance in the central cluster regions \citep[e.g.,][]{elbaz2007}.
The active galactic nuclei (AGN) also tend to be more enhanced and numerous in massive protoclusters at redshifts around 2 to 3, compared to the general field environment or lower redshift clusters \citep[e.g.,][]{pentericci2002,lehmer2009,digby-north2010}. This increased AGN activity appears to be correlated to higher rates of star formation among galaxies in these protoclusters \citep[e.g.,][]{Chapman09,brodwin013,Casey15,gilli2019}.

\begin{figure*}
    \hspace*{-1.2cm}
\centering   
\includegraphics[width=0.89\linewidth]{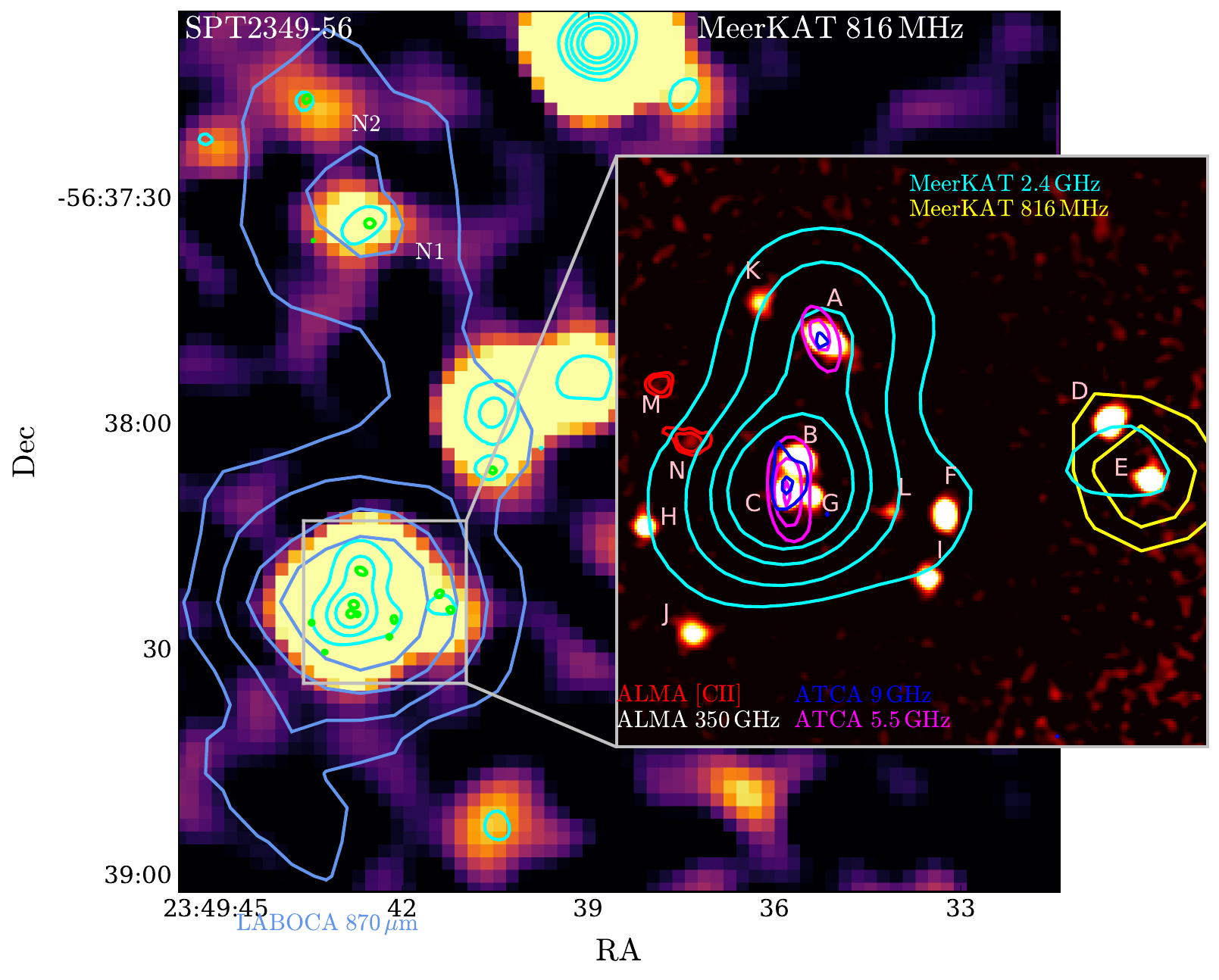}
    \caption{
{\bf Background:} 
MeerKAT 816\,MHz {\tt robust=-1.2} imaging of the \spt\ region, with blue contours highlighting the 110\,mJy extended LABOCA source at 870$\mu$m \citep{Miller}, with ALMA 350\,GHz sources shown (green contours).
%
{\bf Inset:} 
A $22''\times22''$ zoom-in of 
{ALMA} 350\,GHz continuum imaging 
\citep{Hill20} with overlays of MeerKAT  2.4\,GHz (cyan), ATCA 5.5\,GHz (magenta) and 9\,GHz (blue). The point source removed UHF 816\,MHz observation (highlighting source E) is shown as yellow contours.
ALMA sources are named from \citet{Miller} in order of their 850\,$\mu$m flux density. 
For the faintest ALMA sources M,N, we  show [CII] contours (red).
}
    \label{fig:field}
\end{figure*}

At low-to-moderate redshifts ($z<1$),  
both AGN and star formation are largely suppressed in cluster galaxies \citep{Ehlert14,Rasmussen12,vanbreukelen2009,Martini06,Kauffmann04}.
Both mechanical and radiative feedback mechanisms are invoked to accomplish this suppression, which  are naturally produced by AGN. Supporting this, extended X-ray emission observed in some clusters reveal cavities or empty regions in the hot gas, which are consistent with shocked  gas caused by AGN feedback \cite[e.g.][]{fabian2012}.

In the high redshift Universe, however,  galaxy overdensities  
have abundant reservoirs of cold gas to supply star formation, and the ongoing mergers between galaxies expected in the dense environments provide triggers for star formation.
Mergers can also provide the tidal torques necessary for the gas to overcome its angular momentum and fall to the accretion disk of the supermassive black hole (SMBH). 
%
%
Examples at high redshift abound. %
In the MQN01 protocluster at $z=3.25$, the AGN fraction from X-ray is significantly higher ($>$20\%) than in the field and increases with stellar masses, reaching a value of 100\% for $M^* > 3.2\times10^{10}$\,M$_\odot$ \citep{travascio25}.
  The $z\,{=}\,3.09$ SSA22 protocluster identifies 50\% of the SMGs hosting X-ray-luminous AGN \citep{Umehata19} 
-- a clear excess over the 15\% found for field SMGs \citep[e.g.][]{wang2013}.
%
At larger distances, an overdensity of ten SMGs found by the {\it Hershel Space Telescope} at $z\,{=}\,4.0$ \citep{Oteo18} has been studied by {\it Chandra} in the X-ray \citep{vito2020} 
and  in the radio \citep{Oteo18}, revealing no significant excess of AGN activity in the system over field SMGs (22\% versus 15\%, respectively).
Similarly, in the \spt\ protocluster at $z=4.3$, two X-ray AGN were identified \citep{vito24}, with one of them also being radio loud \citep{chapman24}, nominally an even lower AGN fraction in the SMGs ($\sim10$\%).

The X-ray properties observed in the intracluster medium (ICM) of nearby galaxy clusters often do not match what we would expect based solely on gravitational collapse. This suggests that additional sources of energy are at play, such as  
star-forming galaxies and AGN releasing energy into the ICM through outflows, to help to explain these observations \citep{tozzi01,ponman99}.
%
In cosmological simulations, this extra energy is often incorporated by tuning the effects of galactic winds driven by supernovae or AGN activity to match the properties we see locally (e.g., \citealt{lebrun2014,pike2014}). The most successful models suggest that most of this heating occurs early in the Universe's history, but many details remain uncertain based on observations. For example, the details are still debated about when this heating happens, how long it lasts, what the main energy sources are, and how the energy is transferred to the ICM 
\citep[e.g.,][]{mcnamara07,fabian2012}.
%
If this process was common at high redshifts, then energetic events like radio-AGN activity could have played a key role in determining the final energy state of galaxy clusters we observe today. 

Recently, \cite{zhou2026}  identified an immense reservoir of thermal energy in \spt\ through a detection of a strong Sunyaev-Zeldovich (SZ) decrement. 
In energetic terms, the signal is so strong that likely only radio-AGN can provide the required feedback energy input through radio jet mechanical energy, heating an implied hot ICM component of the cluster gas to temperatures exceeding 1.7$\times10^7$\,K. 
This result is supported by a massive excess of cold molecular gas found in \spt, not obviously associated to the individual galaxies \citep{zhou2025}. This cold gas was hypothesized to be a nascent ICM component, revealing that a massive body of gas is presently being heated by both AGN and gravitational collapse. 


\begin{table*}
 \centering
  \caption{Radio-detected sources at $z=4.30$ in \spt.
  Five sources are individually detected by MeerKAT. 
  }
\label{table:radio_full}
\begin{tabular}{lccccccccc}
\hline
ID & RA & Dec & S$_{\rm 816}$ & S$_{\rm 2.4}$  & S$_{\rm 5.5}$ & S$_{\rm 9.0}$  & $\alpha$  & L$_{\rm 1.4, rest}$ & P$_{\rm kin}$\\ 
{} & {} & {} & ($\mu$Jy) & ($\mu$Jy) & ($\mu$Jy) & ($\mu$Jy)   & {} & ($10^{32}$ erg\,$s^{-1}$\,Hz$^{-1}$) & $\times10^{45}{\rm\,erg\,s^{-1}}(f_{\rm cav}/4)$ \\
\hline
A &  23:49:42.68   &  $-$56:38:19.2 & 164$\pm$24 &  71$\pm$9 &  27$\pm$5& 11$\pm$3 &      $-$1.02$\pm$0.10 & 2.5$\pm$0.3 & 0.27$\pm$0.05\\ 
C & 23:49:42.84   &  $-$56:38:25.1  & 718$\pm$25 &  198$\pm$9 &  60$\pm$6&  25$\pm$4&   $-$1.45$\pm$0.08 & 20.1$\pm$0.3 & 2.2$\pm$0.1\\ 
E$^\dagger$ & 23:49:41.22  &  $-$56:38:24.6 & 76$\pm$17 &  22$\pm$7 &  $<12$  & $<13$&   $-$1.16$\pm$0.29 & 1.0$\pm$0.4 & 0.11$\pm$0.04\\ 
N1 & 23:49:42.53   &  $-$56:37:33.2  & 69$\pm$8 &  26$\pm$5 &  $<15$  & $<19$&   $-$0.97$\pm$0.18 & 0.7$\pm$0.2& -- \\ 
N2 & 23:49:43.54  &   $-$56:37:16.6 & 43$\pm$8 &  20$\pm$5 &  $<17$  & $<23$&   $-$0.74$\pm$0.20 & 0.3$\pm$0.1 & -- \\ 
\hline
\hline
\end{tabular}
\\
\flushleft{
$\dagger$ For the 2.4\,GHz measurement in E, the total flux extracted as discussed in Appendix~B is 35$\pm$12\,$\mu$Jy. Here we conservatively measure the radio flux directly at the position of E. The 816\,MHz flux is measured in the {\tt robust=-2} image as shown in Fig.~6. A residual flux measurement of 64$\pm$11\,$\mu$Jy is also derived for E (after subtracting the bright source C) from the {\tt robust=-1.2} image (shown in Fig.~1). \\
}
\end{table*}

This paper presents a search for additional radio detections of members of the \spt\ cluster, with new S-band MeerKAT data, further analysis of the UHF MeerKAT data from \citet{chapman24}, and new deep 5.5\,GHz and 9\,GHz ATCA data.
Section~2 
describes the radio observations. Section~3 
presents the results derived from the source extraction and analysis. In Section~4 
we discuss the overdensity of radio sources and the implications for cluster evolution. 
Throughout our analysis, a Hubble constant of $H_0\,{=}\,70$\ km\,s$^{-1}$\,Mpc$^{-1}$ and density parameters of $\Omega_{\Lambda}\,{=}\,0.7$ and $\Omega_{\rm m}\,{=}\,0.3$ are assumed, resulting in a proper angular scale of 6.88\,kpc/$''$ at $z=4.3$.

\section{Data} \label{sec:data}

\begin{figure*}
    \centering
   \includegraphics[width=.327\linewidth]{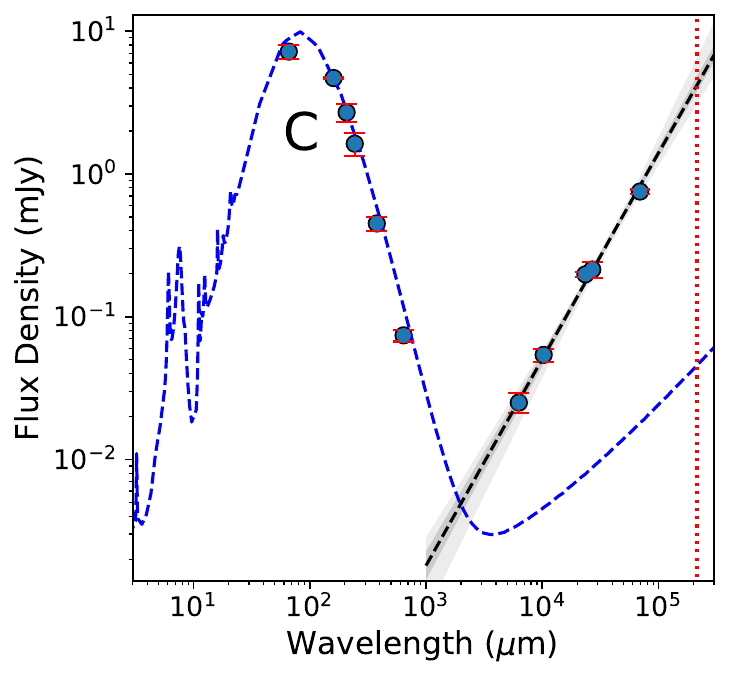}
   \includegraphics[width=.327\linewidth]{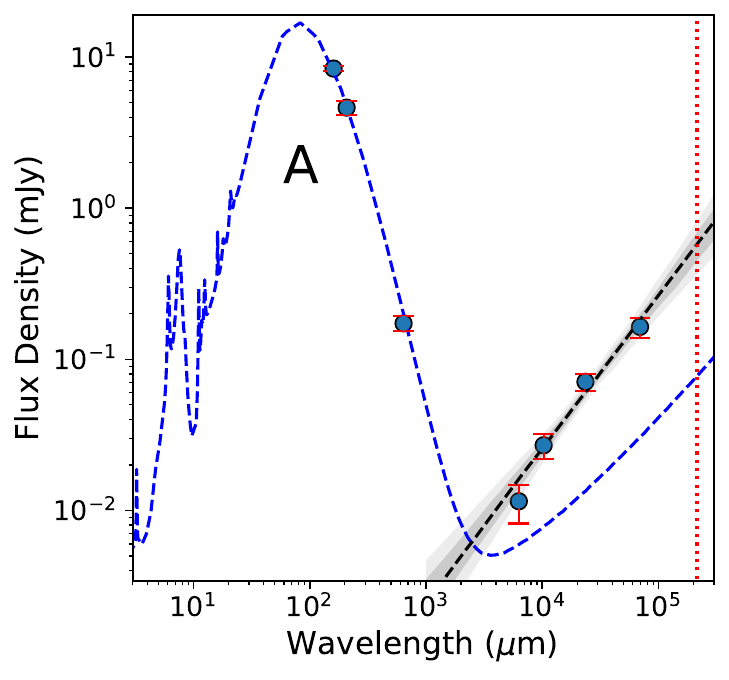}
   \includegraphics[width=.327\linewidth]{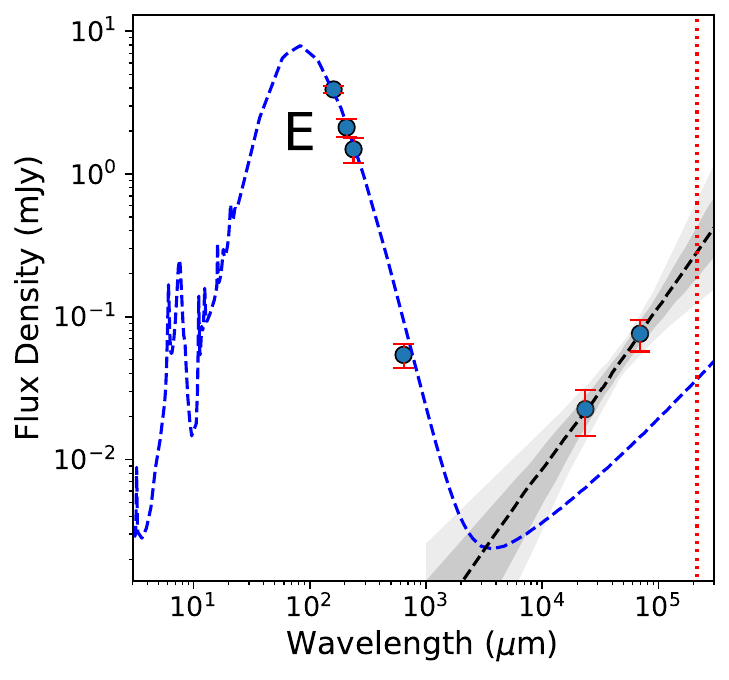}
   \includegraphics[width=.327\linewidth]{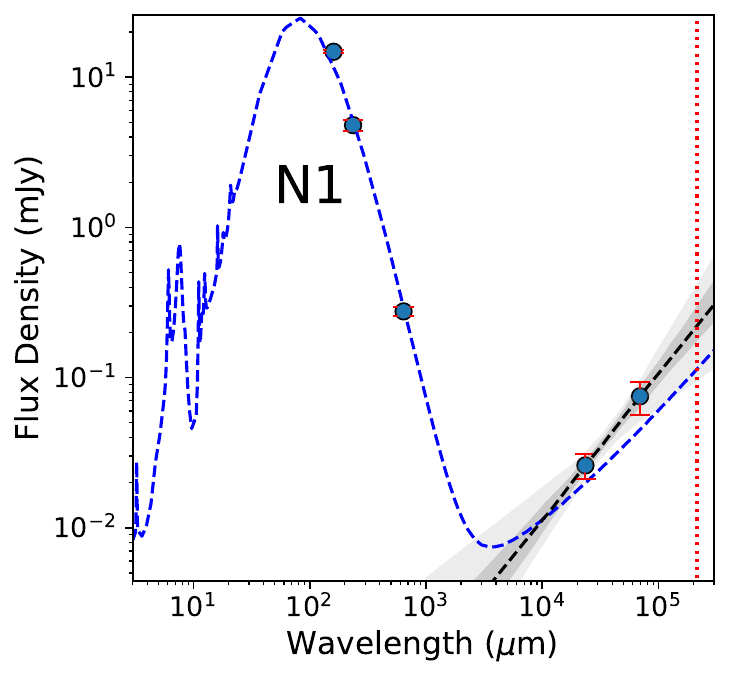}
   \includegraphics[width=.327\linewidth]{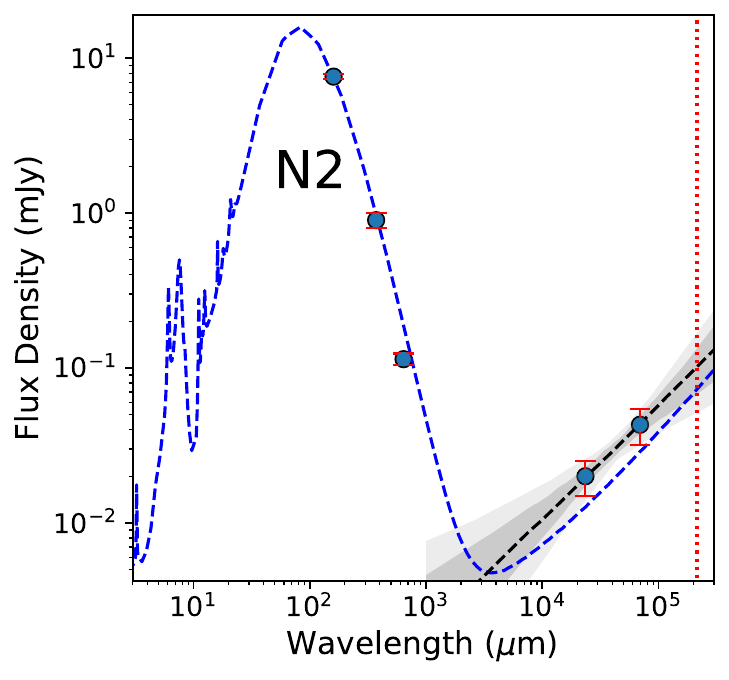}
    \caption{ 
Spectral energy distributions (SEDs) in the restframe wavelength for the five individually-detected sources from MeerKAT. 
Also shown are the ALMA flux densities from \cite{Hill20}, and for 240\,GHz from this work (Appendix~A). For source C, a new 350$\mu$m measurement from \cite{harrington25} is included, suggesting a slightly cooler SED than Arp220 is likely. 
Arp220 lies near the mean of the local radio-far-IR relation. Normalizing Arp220 to the submm photometry reveals that three of the sources show significant excess in radio above the radio-far-IR correlation for star-forming galaxies, while two (N1, N2) show no significant excess over that expected from SF. Sources A and C are detected in the X-ray \citep{vito24}.
The grey shadings show the 1 and 2$\sigma$ uncertainties in the radio spectral index fit (Appendix~B). 
The red line indicates where restframe L$_{1.4}$ is estimated for Figure~3 from Equation~1.
    }
    \label{fig:sed}
\end{figure*}


\subsection{MeerKAT observations and data processing}

\subsubsection{MeerKAT 2.4 GHz data}

 Observations with the MeerKAT radio telescope \citep{Jonas2016} were carried out on \spt\ in the S-band, as part of a survey of eight $z>4$ protocluster fields selected by SPT (Program ID SCI-20230907-SC-01, PI:Chapman).
 
 The S-band covers a usable range of 1.96--3.50 GHz and we selected the S1 sub-band ranging between 1968.75 - 2843.75~MHz. The 875~MHz of digitised bandwidth is split into 4,096 channels, each 214 kHz wide and a correlator dump time of 8\,sec. The observations were carried on 21 December 2023 with a total of 59 of the 64 antennas participating. A total of six 19.5-min scans (1.95~hr on-source) were carried out, interleaved with other protocluster sources to maximise the {\sl uv}-coverage and achieved angular resolution on \spt. A single pointing was centred on the target at ($\alpha, \delta$) = 23h49m42.5s, $-$56d38m22.5s. The absolute flux and bandpass calibrator was J\,0408$-$6545. Time-varying complex gains were solved for using interleaved observations of the gain calibrator J\,0010$-$4153, with an approximate scan length of 1.7\,min every 21\,min. Three 10-min scans were carried out on the flux and bandpass calibrators approximately every 4 hours.

The data were reduced using \textsc{Oxkat}\footnote{https://github.com/IanHeywood/oxkat}, a semi-automated MeerKAT data analysis pipeline fully described in \citet{Heywood2020} and \citet{Heywood2022}, using a similar strategy to that described in \cite{chapman24}. Briefly, the calibration strategy begins with cross-calibration, which includes frequency averaging to a channel width of 0.214~MHz, manual flagging of bad data and automated flagging of known and low-level radio frequency interference (RFI).  Absolute flux calibration assumes J\,0408$-$6545 is a point source with Stokes I flux density of $S_\nu =  8.244$\,Jy\,beam$^{-1}$ with a spectral index of $\alpha  = -1.138$ (where $S_\nu \propto \nu^\alpha$) at a reference frequency of 2406\,MHz.  Bandpass calibration also uses J\,0408$-$6545 as a reference source, while time-variable complex gains are solved for using J\,0010$-$4153 in eight spectral bins and a solution interval of 1.7\,min. The solutions are then all transferred to the target visibilities. Apart from flagging, the cross-calibration process uses standard {\sc Casa} tasks \citep{McMullin2007} and is iterative, with each round applying updated data flags to improve RFI excision and the resultant calibration and image quality. 

 \begin{figure*}
    \centering
   \includegraphics[width=0.495\linewidth]{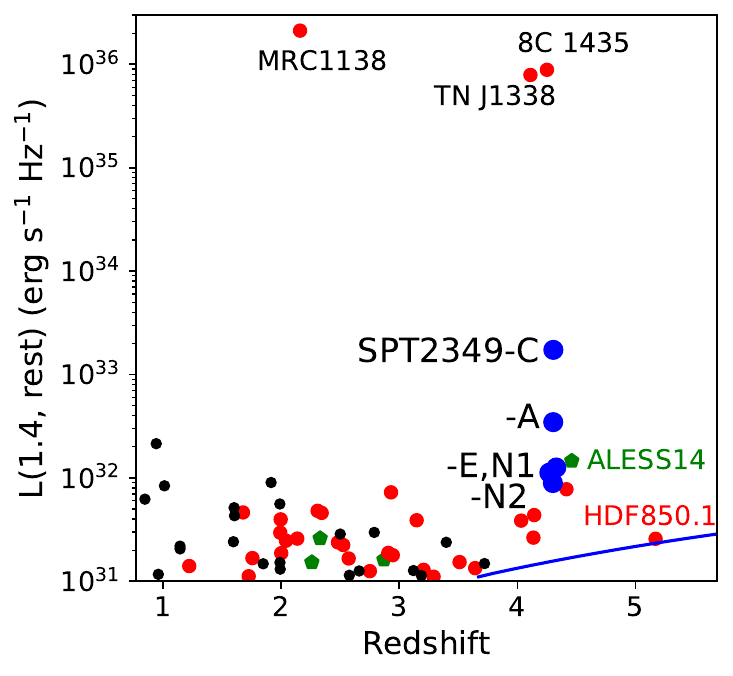}
   \includegraphics[width=0.495\linewidth]{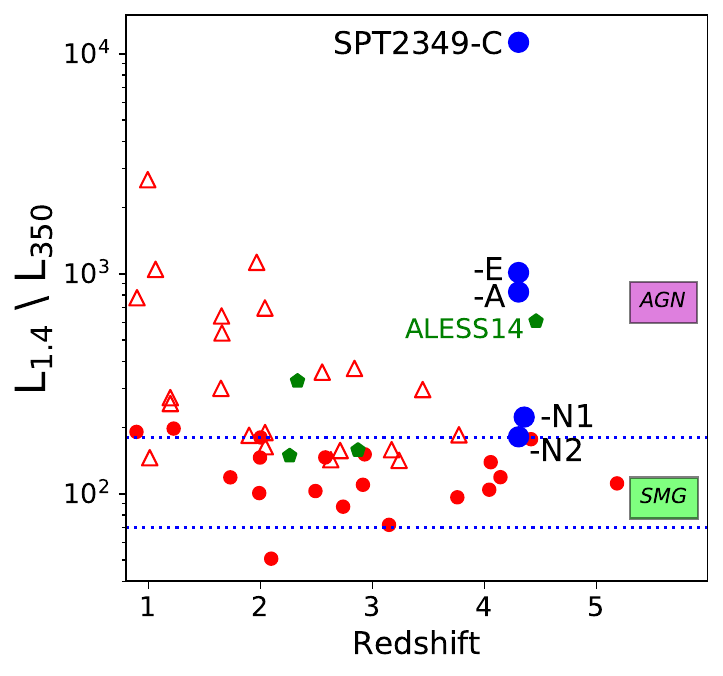}
    \caption{ 
  {\bf Left:} Redshift vs.\ 1.4\,GHz radio power using the GOODS-N sample \citep{barger2017} and radio-excess candidates from the ALESS sample \citep{thomson2014} (green pentagons).  Blue symbols represent \spt\ radio detections (offset in redshift for clarity). Other symbols are as follows: red circles show sources detected above the 3$\sigma$ level at 850\,$\mu$m, while black circles show sources not detected at this level.
    \spt\ source C has about 10 times more radio power than any radio source found in GOODS-N, while A, E, and N1 are amongst the top few percent in radio power found in GOODS-N. The blue line marks the radio flux limit in GOODS-N.
{\bf Right:} Rest-frame 1.4\,GHz luminosity over 350\,GHz luminosity vs.\ redshift, with L$_{350}$ derived using the Arp220 SED. Source C is about 125 times higher than the median relation at rest 1.4\,GHz, with sources A and E being about 10 times higher. N1 and N2 do not appreciably distinguish themselves as expected. For comparison we show submm sources with spectroscopic redshifts in GOODS-N (red circles), and lower limits on radio sources undetected in the submm (red triangles).  The blue dashed line region shows where the submillimeter luminosities and radio luminosities produce consistent estimates of SFRs. None of the GOODS-N SMGs show any excess radio emission over the FIR-radio relation. 
    }
    \label{fig:distribution}
{ \ }\\
\end{figure*}

 We employ a standard approach to self-calibration of MeerKAT data, where an initial deconvolution mask of the target field is created using a shallow unmasked clean, using a {\tt Briggs robust} weighting of -0.3. This mask is then used to create an initial sky model, which is iteratively improved along with the time-variable complex gain solutions through two rounds of antenna-based delay self-calibration using the {\sc Cubical} package \citep{Kenyon2018}, splitting the data into eight spectral bins and time intervals of 64 sec, with multi-frequency synthesis imaging performed during this process using {\sc WSClean} \citep{Offringa2014}. The primary beam's main lobe has a full-width half-maximum of 36~arcmin at 2.4\,GHz, well beyond the protocluster's few arcmin extent. We do not perform any direction-dependent calibration as the S-band residual image has a Gaussian-like noise distribution. As a final step, we use \textsc{DDFacet} \citep{Tasse_2023dd} to generate a $10,125 \times 10,125$ pixel image (pixel size = 0.61$''$, image area $1.7 \times 1.7$\,deg$^2$) with Briggs' {\tt robust} weighting of 0 as a trade-off between desired angular resolution and achieved sensitivity. The resultant point spread function FWHM dimensions are  4.4$''$ by 3.4$''$, at a position angle of -38.8 deg east of north. The rms sensitivity in the vicinity of the target source of this final map is $\sigma = 4.3 \,\mu$Jy\,beam$^{-1}$ and the effective frequency is 2.4\,GHz, since the target is at the field centre.

%


\subsubsection{MeerKAT 816~MHz data}

The MeerKAT 816~MHz data processing and imaging was described in \citet{chapman24} and is not repeated here. The only difference is that the data were re-imaged with modified visibility weighting to better suit comparison with the new MeerKAT S-band data. While MeerKAT has baseline lengths that extend to 8\,km, approximately 70\,\% of its collecting area is concentrated within a central $\sim$1\,km core. This results in a significant range of achieved sensitivity and angular resolution between uniform ({\tt robust = -2}) and natural ({\tt robust = +2}) imaging weightings. To facilitate comparison with the S-band, we produce two UHF images to aid our analysis. We use a {\tt robust = -2} weighting to maximise angular resolution ($\sim$6"), while a second image with {\tt robust = -1.2} (similar to \citealt{chapman24}) has degraded resolution ($\sim$7"), however, this provides higher sensitivity to probe some of the fainter protocluster members (see Fig.~\ref{fig:field} and \ref{fig:radiosub}).  

\subsection{ATCA observations}

\spt\ was observed by the ATCA 
at 5.5\,GHz and 9.0\,GHz between  July 23 to 27, 2023 (PI: Aravena). We used the Compact Array Broadband Backend (CABB) configured in the 1M-0.5k mode, which leads to a bandwidth of 2\,GHz per correlator window with 1\,MHz per channel of spectral resolution. The observations were performed in the most extended ATCA configuration, 6A, with six working 22\,m antennas. The on-source time was 7.5\,hrs in each band. 

The data were edited, calibrated, and imaged using the {\tt Miriad} and {\tt CASA} packages. Data affected by known radio frequency interference (RFI) or with bad visibility ranges were flagged accordingly. We estimate an absolute calibration uncertainty of $5\%$ at 5.5\,GHz and 10\% at 9.0\,GHz.  We inverted the visibilities using natural weighting, leading to beam sizes of $2.7^{\prime\prime}\,{\times}\,1.5^{\prime\prime}$ and $2.1^{\prime\prime}\,{\times}\,1.2^{\prime\prime}$ at 5.5 and  9.0\,GHz, respectively, with associated RMS noise values  of  3.6 and 2.9\,$\mu$Jy\,beam$^{-1}$.


\section{Results}\label{sec:results}

Figure~1 shows the MeerKAT UHF 816\,MHz imaging of the \spt\ region from the new analysis presented in this work, highlighting the 110\,mJy extended 870$\mu$m APEX-LABOCA source which defines the active protocluster region. The new S-band 2.4\,GHz map (shown contoured over the UHF image) cleanly separates the radio emission from five known cluster sources: C, A, E, N1, and N2, where the two northern sources are also well detected individually in the UHF map.  Fluxes are listed in Table \ref{table:radio_full}.
%

Figure~\ref{fig:field} also displays the ATCA\,5.5 and 9\,GHz maps, revealing two well-detected (17 and 8$\sigma$) sources  near the core of \spt\ corresponding to ALMA sources C and A. Both show signs of extension beyond the beam, with observed sizes {derived from 2D Gaussian fitting} of $4.2''\times1.6''$ (PA=4.7$^\circ$) and $3.5''\times1.6''$ (PA=12.1$^\circ$) respectively.
Deconvolved from the beam, the true sizes are $3.2''\pm1.1''$ for C and $2.2''\pm1.3''$ for A in roughly the north-south direction (and unresolved in the orthogonal direction).
At 9\,GHz, source C is unresolved while source A is only  marginally detected. 
Sources E, N1 and N2, all undetected by ATCA, lie at 12$''$, 52$''$ and 69$''$ respectively from the phase center, resulting in attenuation of $\sim$25\% for the northern sources at 5.5\,GHz, and 50\% at 9\,GHz.

\subsection{Source C: revised radio power and spectral index}

Source C was previously identified in \cite{chapman24} as the only radio source in \spt, with a radio luminosity sufficiently high as to be able to significantly heat any nascent ICM gas present in the core. 
With the new MeerKAT data, we are able to refine the analysis, noting that the radio luminosity for source C was previously overestimated by $\approx$10\% \citep{chapman24}, as the blended radio emission with source A at $\sim$800\,MHz from MeerKAT and ASKAP was partially ascribed to C. 
Here we cleanly separate the emission from A and C using the 
816\,MHz MeerKAT data ($\sec$~2.1.2) by reimaging with Briggs {\tt robust}$-2$, and deconvolving the emission with a double point source model (appendix~B). 
The revised MeerKAT 816\,MHz flux, together with the three higher frequency bands,  constrains a  slope with high significance,
$\alpha=-1.45\pm0.08$ (Figure~\ref{fig:sed}), essentially unchanged from the previous estimate ($\alpha=-1.45\pm0.16$).
We compute the rest-frame radio luminosity using the equation
\begin{equation} 
L_{1.4}=\left( 4\pi d_{L}^2\, S_{1.4} / 10^{29} \right) (1+z)^{-(\alpha+1)}\, {\rm erg\,s^{-1}\,Hz^{-1}} 
\end{equation} 
where $d_{L}$ is the luminosity distance (in cm) and $S_{1.4}$ is the  flux density in units of $\mu$Jy measured at 1.4\,GHz, rest, with  $S_\nu\,{\propto}\,\nu^{\alpha}$. At the \spt\ redshift of 4.3, $L_{1.4}$ corresponds to an observed frequency of 264\,MHz or wavelength of 214,287\,$\mu$m as indicated on Figure~\ref{fig:sed}.
Source C accordingly has a revised 10\% lower radio luminosity (L$_{1.4, \rm rest}$ = $2\times10^{33}$\,erg s$^{-1}$ Hz$^{-1}$). However this still represents an excess of 125 times above the radio-far-infrared (far-IR) relation at this frequency for star-forming galaxies \citep{helou1985}.

We also characterize the radio morphology at 5.5\,GHz using the new high-resolution ATCA imaging, finding a significant extent centered about C of $3.2''\pm1.1''$, suggesting a projected bipolar radio jet is emanating from the AGN with a length of at least 22\,kpc (see $\sec~4$). This interpretation is strengthened by this candidate radio jet being oriented at 5$^\circ$ (East of North) almost perpendicular to the [CII] disk major axis which lies at a PA=125$^\circ$  
\citep{chapman24}.  The sensitivity of the ATCA 5.5\,GHz imaging would only be able to detect the brightest parts of the jet (by analogy to lower-redshift and/or more powerful radio jets studied at high resolution), and the extent of the radio jets influence into the cluster potential could be substantially larger.
The extent of the jet is not significantly detected at 9\,GHz despite the improved spatial resolution achieved over 5.5\,GHz, however the emission at 9\,GHz is also well localized to ALMA source C. 
While the 5.5\,GHz emission does extend partially over ALMA source B, the body of evidence above (along with C being an X-ray AGN) suggests this is indeed the radio jet from C projected over source B, rather than B contributing to the radio luminosity. Lower level radio emission in B is expected from its far-IR luminosity, but this would be a factor $\sim$20$\times$ fainter than observed.



\subsection{Two new radio AGN in \spt} 

Four additional cluster members are individually detected by MeerKAT in S-band and the new UHF-band weightings (\S~2.1.2), namely sources A and E from the original \cite{Miller} sample, along with sources N1 and N2 in the northern sub-group from \cite{Hill20}. The flux measurements are described in Appendix~B.
Sources A and E  stand out at 816\,MHz and 2.4\,GHz as showing significant excess radio emission above the expectation from star formation alone, along with steep radio spectra characteristic of AGN (figures~2 and 3). 
Their properties are summarized in table~\ref{table:radio_full}, while figure~\ref{fig:distribution} compares these sources' radio luminosities and radio excess to literature samples of  radio sources and SMGs.
Radio-loud AGN are typically suggested as having $\sim$5$\times$ excess above the radio-far-IR relation \citep[e.g.,][]{daddi2017,gentile2025}, which is easily satisfied by all of C, A, and E (Figure~\ref{fig:distribution}). While N1 does not exhibit radio excess, a possible AGN indication is present in its strong high-J CO emission (Appendix~A), as well as its  low [CII]/FIR ratio \citep{Hill20}. 


Source A was previously identified as an X-ray-luminous AGN \citep{vito24}, and  MeerKAT and ATCA observations reveal it to also be a relatively steep spectrum ($\alpha=-1.0\pm0.1$) radio-loud AGN, with a radio excess of nine times above that implied by star formation.  Its radio power is $\approx$8$\times$ lower than source C at 1.4\,GHz (rest).
Source A is detected with ATCA at 5.5\,GHz and 9\,GHz, and as with source C shows evidence for some radio extent beyond the beam ($2.2''\pm1.3''$ or 15\,kpc in length), possibly also implying jet emission. As with source C, this interpretation is strengthened by the somewhat opposing orientations of the jet (PA=12$^\circ$) and the [CII] disk (PA=54$^\circ$) -- a difference which could even be interpreted as bipolar if the disk were sufficiently tilted into the line of sight. 
The radio spectrum is apparently much steeper at higher frequencies with ATCA than implied by MeerKAT at lower frequencies. This turnover may represent a very young radio source (\S~4.2). However, there is also a possibility that the radio source has a large spatial extent and flux loss at higher frequencies may affect the spectral slope.

Source E is steeper in radio slope than A  ($\alpha=-1.2\pm0.3$), although less well constrained as it is undetected at 5.5\,GHz, and has more uncertainty in the MeerKAT fluxes (Appdenix~B). It represents a lower radio power source, but still with a radio excess of 9.4 times that inferred from its SFR. 
 
The N1 and N2 sources are identified by MeerKAT as not having a 
signficant radio excess over that expected from the radio-far-IR correlation.
However N1 is also revealed to have a strong and compact $^{12}$CO(11-10) line emission (Appendix A). N1 is the only source in \spt\  showing a spectral line energy density (SLED) more like the AGN Mrk231 than the starburst galaxy M82. Source N1 is not detected in the X-ray \citep{vito24}, indicating either a deeply obscured Compton-thick AGN (the rest frame optical counterpart shows very large extinction -- \citealt{hill2022}), or a modest mass SMBH below that expected from its stellar mass \citep{hill2022}. Another explanation could be sub-Eddington accretion.
N1 is an extremely luminous SMG, with a far-IR luminosity over $3\times10^{13}$\,L$_\odot$ \citep{zhou2025,Hill20}, with an implied SFR almost 3000\,M$_\odot$ yr$^{-1}$. As such the copious radio emission associated with its SFR could  dominate over a weaker radio AGN component. 

\subsection{Radio contribution from star formation}

Sources N1 and N2 are the most far-IR-luminous sources in the cluster, with N2 being comparable to the brightest source A in the core region.  Both are individually detected in the MeerKAT UHF and S-band (Table~1), revealing the radio emission and  synchrotron slope ($\alpha\sim-0.7$) expected from the correlation with far-IR luminosity (Figure~\ref{fig:sed}).  
While both sources exhibit a small excess above the median  in the local radio-far-IR relation, it is consistent within 2$\sigma$ of the radio photometry errors and uncertainties in the slope $\alpha$ (Appendix~B.1), and well within the 0.2~dex scatter in the local relation. 
As such there is not any strong evidence for an increase in radio-far-IR claimed at high redshifts (e.g., \citealt{gentile2025}).

The detections of N1 and N2 suggest that marginally significant radio emission from some of the other brighter cluster members may be blended with the central radio AGN, as hinted at in the S-band image in Figure~\ref{fig:field}.
In the residual map of the core at S-band, after subtracting sources C, A, and E, there is a marginal excess in radio emission revealed in the region of sources F and D, which would be consistent with that expected ($\sim5\mu$Jy) from the radio-far-IR relation. The residual emission does not however represent even a $>3\sigma$ detection.






\section{Discussion}\label{sec:discuss}

\subsection{Overdensity of radio-AGN}
Three radio-AGN in this small $z=4.3$ core volume (the central 130\,kpc) is an enormous overdensity, a factor of $>10^7\times$ larger than found in deep radio surveys followed up spectroscopically like that in GOODS-N \citep{barger2007} or COSMOS \citep{smolcic2017}. 
%
When we consider that radio-AGN typically represent only $\approx$10\% of the total AGN population \citep{panessa2019}, our finding would naively suggest that there could be substantially more radio-quiet AGN in the \spt\ core. 
In fact, these three central radio-AGN represent all of the AGN currently known in the core region. 

There are $\sim40$ cluster members in the protocluster identified to date, the vast majority selected by submm continuum or [CII] line emission \citep{sulzenauer25}, and further observations (e.g., with JWST) may identify other galaxies as hosting lower-luminosity (radio-quiet) AGN. 
Four AGN amongst the $\sim$30 submm-emitting cluster members now cataloged is not particularly unusual, representing a fairly typical 13\% AGN fraction for protoclusters, although these numbers are sensitive to depths of surveys at different wavelengths. If we consider only the 14 ULIRGs in \spt\ with SFR$>100$\,M$_\odot$\,yr$^{-1}$, the AGN fraction is 29\%.

Other claimed multiple radio-AGN in protoclusters, such as the two identified by \cite{daddi2017} in the $z=2.5$ Cl\,J1001 would not come close to being classified as radio-loud AGN by our criteria (as shown in Figure~\ref{fig:distribution}), with radio luminosities only $\approx5\times10^{31}$ ergs s$^{-1}$ Hz$^{-1}$ at 1.4\,GHz, rest, and modest excess over the radio-far-IR expectation.
Even the closest known analog to \spt\ in the dense DRC protocluster \citep{Oteo18} shows only a single radio-AGN, along with two X-ray identified AGN \citep{vito2020}. Finding a single radio-AGN is typical in general of many protocluster fields, some of which are actually selected using the radio-AGN itself -- e.g., the Spiderweb \citep{dannerbauer2014}. This basic finding is possibly related to a key aspect of galaxy cluster evolution, that radio jets are thought to provide an important feedback mode in galaxy clusters by preventing the cooling of hot (X-ray) gas surrounding central galaxies \citep{McNamara2012}.


\subsection{Energetics of the radio-AGN}

Simple energetic considerations establish that radio jets can, in principle, have drastic effects on the gas-phase baryons in and surrounding massive galaxies. \cite{zhou2026} have argued that the large overdensity of radio-AGN in \spt\ represents an energetically favorable way to explain the energy input into the ICM that may drive the large SZ signal they observe. 
The total energetic output from these three radio-AGN (L$_{1.4}=2.6\times10^{26}$ W Hz$^{-1}$) is not significantly larger than the previous (overestimated) radio luminosity ($2.2\times10^{26}$ W Hz$^{-1}$) presented for the single central AGN \citep{chapman24}. However, the fact that three likely randomly oriented radio jets are impacting the ICM may facilitate seeing such an early, large and uniform tSZ decrement aligned with the protocluster center of mass. This contrasts all other protocluster tSZ detections found at $z\sim2$, where large spatial offsets are observed between the SZ decrement and the core region of the cluster \cite{mantz2018,Gobat2019,mascolo2023}.



\cite{chapman24} have speculated that the radio-AGN in \spt\ can be fueled by hot gas in radio-mode instead of radiative-efficient accretion through recent mergers, which can provide strong kinetic feedback with substantial energy injection on the nascent ICM in the protocluster. 
We use the correlation between the cavity power ($P_{\rm cav}$) and radio luminosity at 1.4\,GHz ($L_{\rm 1.4\,GHz}$) to estimate the kinetic power from radio luminosity alone for each of the radio AGN (listed in Table~\ref{table:radio_full}), which can be described using \cite{heckman14}:
\begin{equation}
    P_{\rm cav}=7\times10^{43}{\rm \,erg/s}\times f_{\rm cav}\left(\frac{L_{\rm 1.4\,GHz}}{10^{25}{\rm W\,Hz^{-1}}}\right)^{0.68} ,
\end{equation}
where the enthalpy factor $f_{\rm cav}=4$ for relativistic plasma. 
Shocks induced by the radio jet can cause additional heating, which could imply a higher $f_{\rm cav}>4$ \citep{Nusser2006}. 
This scaling relation yields a total kinetic power from the three radio-AGN of $\dot{E}_{\rm kin, radio}=(2.6\pm0.3)\times10^{45}{\rm\,erg/s}\times(f_{\rm cav}/4)$. 

Assuming $f_{\rm cav}=4,$ 
we can estimate the energy input: 
$$        \Delta E_{\rm inject} \approx \dot{E}_{\rm kin,radio} t_{\rm radio}$$ 
\begin{equation}
    = \left[(8.7\pm0.8)\times  \left(\frac{t_{\rm radio}}{100\rm\,Myr}\right) 
        \right]
        \times10^{60} \rm\, erg. 
\end{equation}
Adopting $t_{\rm AGN}=100\,\rm Myr$ for  AGN activities, we obtain a total energy injection of $(8.7\pm0.8)\times10^{60}\rm\,erg$ to be stored in the nascent ICM, comparable to that estimated directly from the tSZ decrement ($\sim10^{61}$\,erg) in \cite{zhou2026}. 
In this ``High-Temperature (Feedback)'' scenario, powerful energy injection from the known radio-AGN have super-heated the gas \citep{zhou2026,chapman24}, creating the tSZ signal through a modest ICM gas mass heated to high temperatures ($>3\times10^{7}$\,K). However, the strong tSZ signal requires a nearly 100\% coupling efficiency if the energy is solely provided by the identified radio-loud AGNs.
The energetic jets may not couple efficiently to the bulk of the ICM gas \citep{babul2013,yang2016,cielo2018}.
This scenario is however testable -- 
this same feedback would also suppress the X-ray luminosity expected from the ICM compared to that generated by gravitational energy (e.g., accretion shocks -- \citealt{shi16}) from a massive halo. Such a low-mass ICM heated to a higher temperature will puff up gas and lead to X-ray cavities, which don’t increase the X-ray photons. However the X-ray spectrum would substantially shift to higher energy, which could be detectable in the X-ray, by XMM-{\it Newton} for example.

\subsection{Synchronizing three radio-AGN}
A key question is how the three radio-AGN in the core of \spt\ can be simultaneously active in this small volume. 
\cite{wang_gan2008} provide a simplified model of jet power from active galactic nuclei, whereby disk accretion is combined with the sum of two mechanisms of extracting energy magnetically from a black hole accretion disk, i.e., the Blandford-Payne (BP) \citep{blandfordpayne1982} and the Blandford-Znajek (BZ) \citep{Blandford1977} processes. In the BP process, the disk matter is channeled into the outflow/jet by way of the
poloidal magnetic field lines frozen in the disk. The channeled gas is then accelerated via work done by the magnetic torque.
The BZ process, although not physically viable on its own \citep{punsly} provides an  approach used in \cite{wang_gan2008} to explain why some AGN are intensely bright in radio waves (radio-loud) while others are not (radio-quiet). Rapidly spinning SMBHs with sufficient magnetic fields are more likely to produce powerful jets. As the black hole rotates, the magnetic field lines, which are frozen into the plasma of the accretion disk, are twisted and stretched, 
converting the black hole's rotational energy into electromagnetic energy. 
This extracted energy is then channeled along the twisted magnetic field lines and flung outwards as a highly collimated beam of ionized matter, resulting in a relativistic jet.
There are possible mechanisms to keep radio galaxies active longer than  $\sim$100\,Myr \citep{Brienza2017}, but arguing for these exceptional circumstances happening on the scale of the central SMBHs is contrived, even in the overdense environment of \spt. 
A more plausible explanation is simply that the mega-merger \citep{Rennehan} resulting from the remarkably overdense core environment (more than any ever identified in the Universe) has driven all the luminous AGN into a radio-loud mode within the roughly 100\,Myr timescale \citep[e.g.,][]{heckman24}. 


\subsection{Radio source properties and lifetimes}

Two new pieces of information have emerged from our updated study, resolved jet sizes and steep spectral indices ($\alpha$) confirmed at higher frequencies ($>5$\,GHz). 
A steep spectrum is typically derived from self-absorbed synchrotron, along with a lack of electron injection (e.g., \citealt{radcliffe2021}).  Thus the steep $\alpha$ measured in these three radio-AGN could represent a dying radio source.  
On the other hand, these properties could be explained as due to a confined or ``frustrated'' radio source inside a dense medium, still building in radio power. This would be consistent with the observed hot ICM in \cite{zhou2026}. 
These confined radio sources are sometimes referred to as compact steep spectrum sources, or CSS \citep{padovani2017}. However, the luminosities of the \spt\ radio sources are low relative to  typical GHz-peaked CSS sources, even in the case of source C. 
While the three radio-AGN in \spt\ span a factor of $\sim20$ in L$_{1.4}$, it is possible that they may have had a more equal or different distribution in radio power in the recent past, since it is unclear what the history or lifetime of any of the three is.
If self-absorbed synchrotron is contributing to the steep spectrum, the observational constraints would mean that the break frequency is below about 5\,GHz in the rest frame (940\,MHz in observed). This break frequency (as seen explicitly in source A) can provide a constraint on the age of the radio source. However if the radio emission is due to a CSS then it would have to be older than 500\,Myr to have a break frequency below 900\,MHz rest (4.8\,GHz observed) \citep{padovani2017}. 

\section{Conclusions}

We have found two additional radio-excess galaxies (A and E) in \spt\ (both $\sim10\times$ above the radio-far-IR correlation at 1.4\,GHz), for a total of three radio-AGN in the cluster with the previously identified source C. Two of these AGN (A and C) are also X-ray detected \citep{vito24}.
One other candidate AGN (source N1) is detected in the radio, but does not show obvious excess over that expected from SF. 
The AGN in N1 is only identified through an excess in CO(11--10) emission (comparable to Mrk231) indicating a flat CO SLED to at least J=11. The lack of enhanced radio or X-ray emission indicates that it is Compton-thick and radio-quiet.  

$\bullet$
Four AGN amongst the $\sim$30 submm-emitting cluster members now cataloged \citep{sulzenauer25}  represent a fairly typical  13\% AGN fraction for $z>2$ protoclusters, although this fraction goes up to 29\% considering only the 14 ULIRGs.

$\bullet$
However, the overdensity of radio AGN in \spt\ is larger than has been previously reported in the literature, although the statistical uncertainly of three versus one is small.

$\bullet$
The three central radio-AGN are likely heating the nascent ICM, providing an estimated power of $(2.6\pm0.3)\times10^{45}{\rm\,erg/s}\times(f_{\rm cav}/4)$, and could give rise to the abnormally large tSZ signal detected in \cite{zhou2026}.

$\bullet$
The direct evidence for compact radio jets from the 5.5\,GHz morphology, and the more tentative evidence for a spectral break below $\sim$5\,GHz (in source A) argue for young radio sources and/or confined medium, again supporting an interaction with the hot ICM detected in this system.

\section*{acknowledgements}

%
The MeerKAT telescope is operated by the South African
Radio Astronomy Observatory, which is a facility of the National Research Foundation, an agency of the Department of Science and Innovation. The research of RPD is supported by the South African Research Chairs Initiative (grant ID 77948) of the Department of Science and Innovation
and National Research Foundation.
The Australia Telescope Compact Array is part of the Australia Telescope National Facility (https://ror.org/05qajvd42), which is funded by the Australian Government for operation as a National Facility managed by CSIRO.
%
%
The National Radio Astronomy Observatory is a facility of the National Science Foundation operated under cooperative agreement by Associated Universities, Inc.
%
This paper makes use of the following ALMA data: ADS/JAO.ALMA\#2015.1.01543.T, ADS/JAO.ALMA\#2018.1.00058.S, and\\ ADS/JAO.ALMA\#2021.1.01010.P. 
ALMA is a partnership of ESO (representing its member states), NSF (USA) and NINS (Japan), together with NRC (Canada), MOST and ASIAA (Taiwan), and KASI (Republic of Korea), in cooperation with the Republic of Chile. The Joint ALMA Observatory is operated by ESO, AUI/NRAO and NAOJ.
S.C.C., D.Z., J.B., C.H., R.H., V.P., and G.W.\ gratefully acknowledge support for this research from NSERC.
This research was supported in part by grant NSF PHY-2309135 to the Kavli Institute for Theoretical Physics (KITP).
Manuel A.\ acknowledges support from FONDECYT grant 1211951, CONICYT + PCI + INSTITUTO MAX PLANCK DE ASTRONOMIA MPG190030. 


\appendix

\section{CO SLED analysis of SPT2349$-$56 members} \label{sec:appendixA}

\subsection{New ALMA observations}

Extensive ALMA properties of these radio-detected \spt\ sources (A, C, E, N1, N2) have already been published \citep{Miller,Hill20,Rotermund21,hughes25,chapman24}.
Here we present  new ALMA observations (Table~\ref{table:lines}), supporting our measurements of line emission in the context of searching for AGN.

The Cycle 8 program (2021.1.01313.S, PI: Canning), described initially in \cite{chapman24} for sources B, C, and G, observed $^{12}$CO(11--10) ($\nu_{\rm rest}\,{=}\,$1267.01\,GHz) and continuum at about 240\,GHz in Band~6.
These observations, carried out on 2022, September 1, used the C-4 array configuration with baselines of 15 to 784\,m, giving a naturally-weighted synthesized beam size of 0.47$^{\prime\prime}$. J2357$-$5311 and J2258$-$2758 were used to calibrate the amplitude, while J2357$-$5311 and J2336$-$5236 were used to calibrate the phase.
All the data were calibrated using the standard observatory-supplied calibration script. Imaging was done using the {\tt CASA} task {\tt tclean}, using Briggs weighting with a robust parameter of 0.5, and in all cases channel widths were averaged down to a common 15.625\,MHz.

Reduction and analysis followed the same procedures described in \cite{hughes25}. 
The same apertures used by \citet{Hill20} to extract line and  continuum measurements were applied to all sources in order to extract one-dimensional CO(11--10) spectra for each source. 
New continuum flux densities and line strengths are listed in Table~\ref{table:lines}, updating those three sources previously presented in \citep{chapman24} and the new continuum measurements are also shown in Fig.~\ref{fig:sed}. 

\begin{figure*}
 \centering  
 \includegraphics[width=0.695\linewidth]{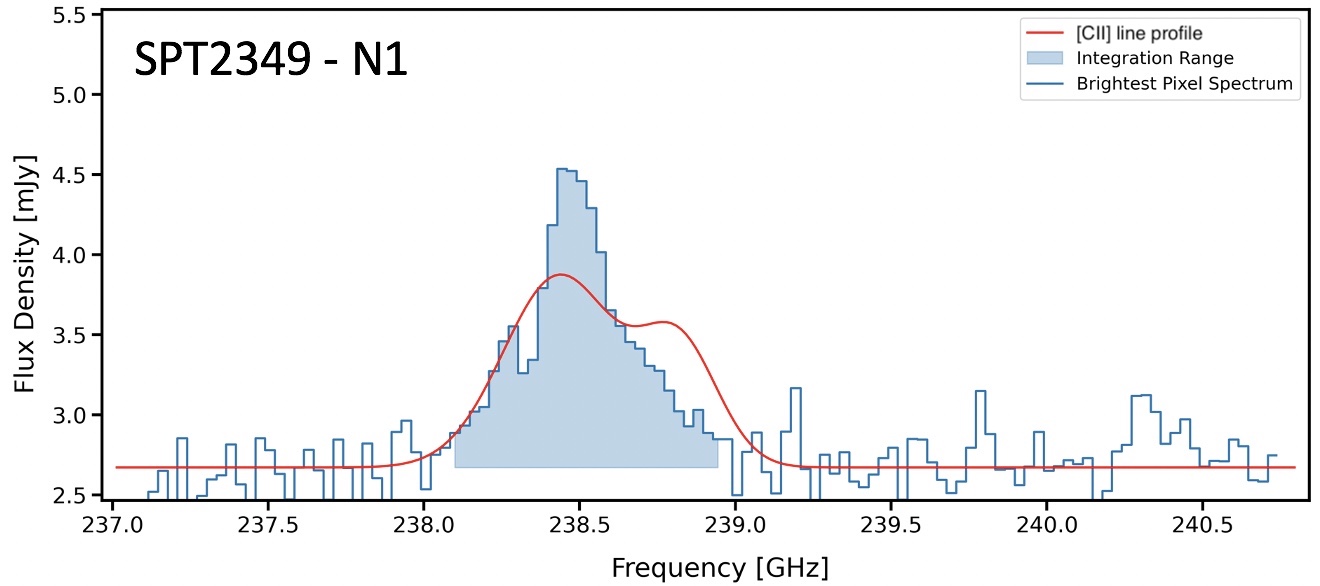}
    \caption{ 
   CO $J_{\rm upper}\,{=}\,$11 spectrum of N1, significantly brighter than any other in the protocluster, and much narrower than the CO(4--3) and [CII] lines. Analysis of the CO SLED suggests a possible AGN-driven XDR contribution to the gas excitation (Figure~\ref{fig:sled}) The red line denotes the aperture-integrated [CII] spectrum of N1.}
    \label{fig:N1co1110}
{ \ }\\
\end{figure*}
  
\begin{figure*}
    \centering
 \includegraphics[width=0.695\linewidth]{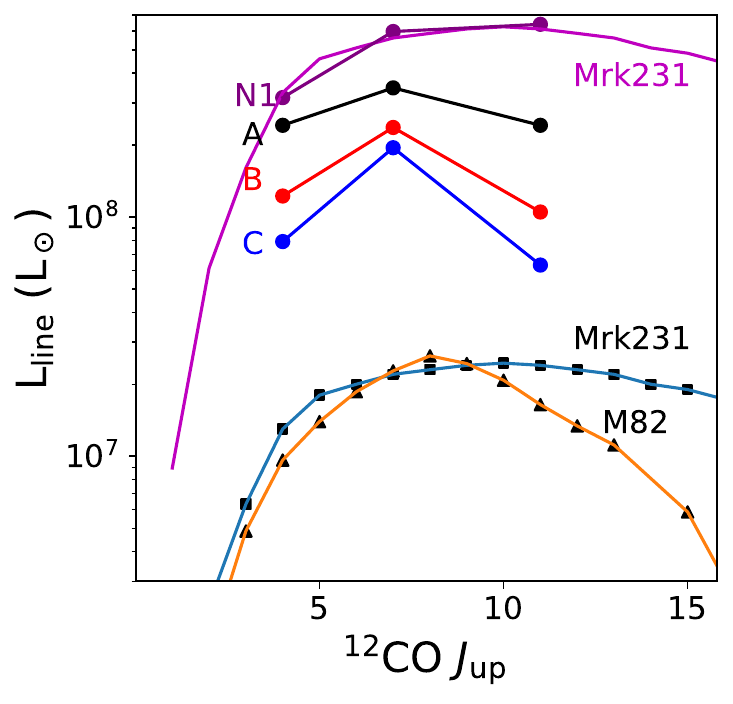}
    \caption{ 
  SLEDs for the sources detected at $>5\sigma$ in the CO $J_{\rm upper}\,{=}\,$11 transition. New spectral data for the $J_{\rm upper}\,{=}\,$11 lines are shown in Figure~\ref{fig:N1co1110}. 
  We compare to the $L_{\rm FIR}\,{=}\,3\,{\times}\,10^{12}$\,L$_\odot$ AGN-dominated galaxy Mrk231 \citep{vanderWerf},
  and the $L_{\rm FIR}\,{=}\,3\,{\times}\,10^{10}$\,L$_\odot$ starburst M82 \citep{kamenetzky2012}, here normalized to Mrk231 at CO(7--6). 
  Source N1 shows an excess J=11 emission at very high significance, with almost perfect agreement with the Mrk231 SLED  modeled with a strong XDR contribution \citep{vanderWerf}. 
  The radio-AGN A and C, show SLEDs more consistent with starburst galaxies. The radio-AGN E is not significantly detected in CO(11--10), but is also consistent with an M82-like excitation.
}
    \label{fig:sled}
{ \ }\\
\end{figure*}

\begin{table*}
 \centering{
  \caption{CO(11--10) measurements and associated 240\,GHz continuum in the \spt\ galaxies with significant ($>5\sigma$) detections.
  }}
\label{table:lines}
\begin{tabular}{lccccc}
\hline
ID & RA & Dec & S$_{\rm 240}$ & S$_{\rm CO(11-10)}$ & L$_{\rm CO(11-10)}$ \\ 
{} & {} & {} & (mJy) & (Jy\,km/s)   & ($10^{8}$ L$_\odot$) \\
\hline
A &  23:49:42.68   &  $-$56:38:19.2 & 2.8$\pm$0.3 &  0.37$\pm$0.06 &     24.2$\pm$4.4\\ 
B &  23:49:42.78 & $-$56:38:23.8 & 2.2$\pm$0.2 &  0.20$\pm$0.03 &  10.6$\pm$2.1\\
C &  23:49:42.84 & $-$56:38:25.1 & 1.9$\pm$0.2 &  0.18$\pm$0.03  &  6.3$\pm$1.2\\ 
N1 &  23:49:42.53 & $-$56:37:33.2 & 4.8$\pm$0.2 &  0.98$\pm$0.08 &     64.0$\pm$4.9\\ 
\hline
\hline
\end{tabular}
\\
\flushleft{
}
\end{table*}

\section{Analysis of the MeerKAT data}

\subsection{Radio flux measurements}
\begin{figure*}
 \centering  
 \includegraphics[width=0.895\linewidth]{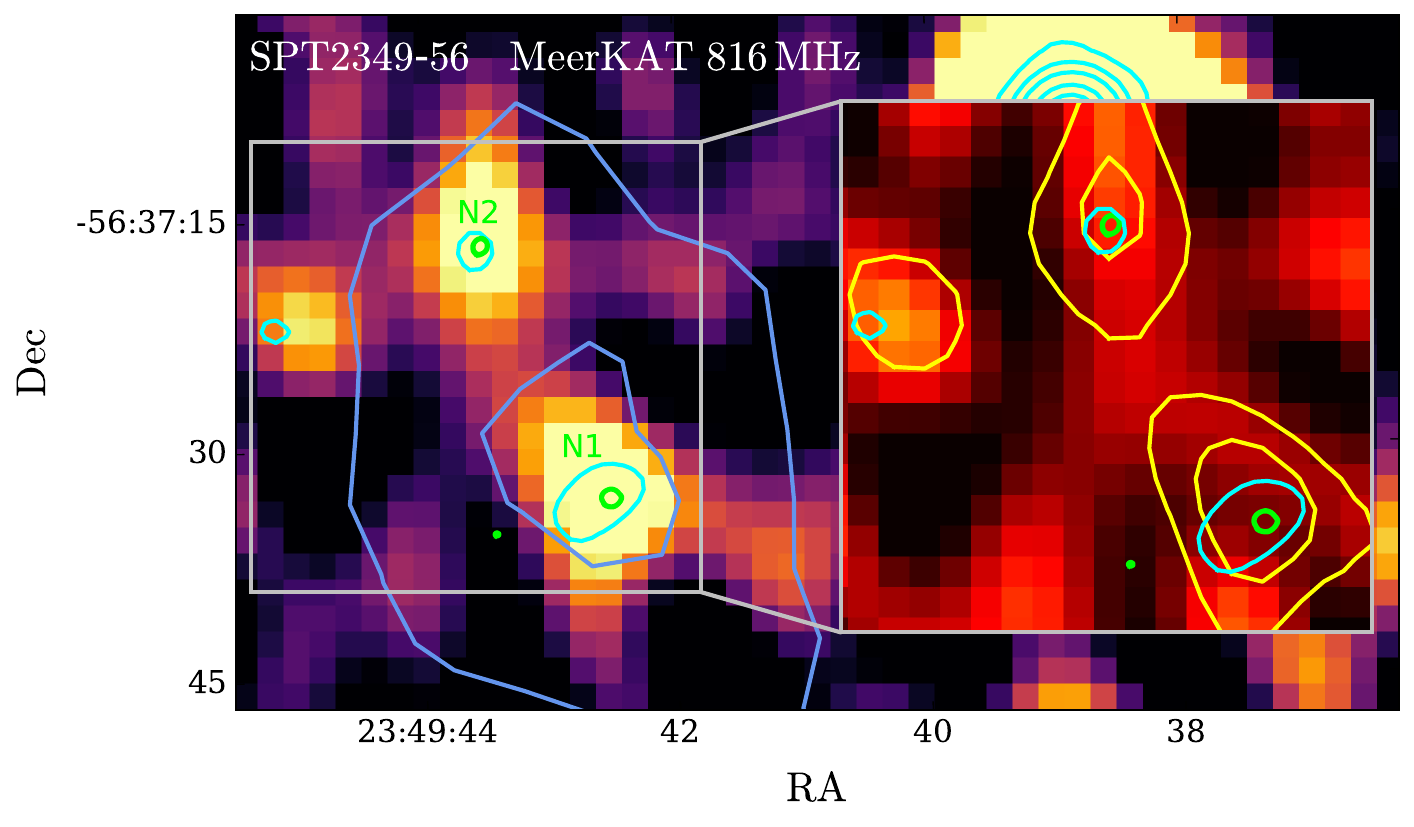}
 \includegraphics[width=0.895\linewidth]{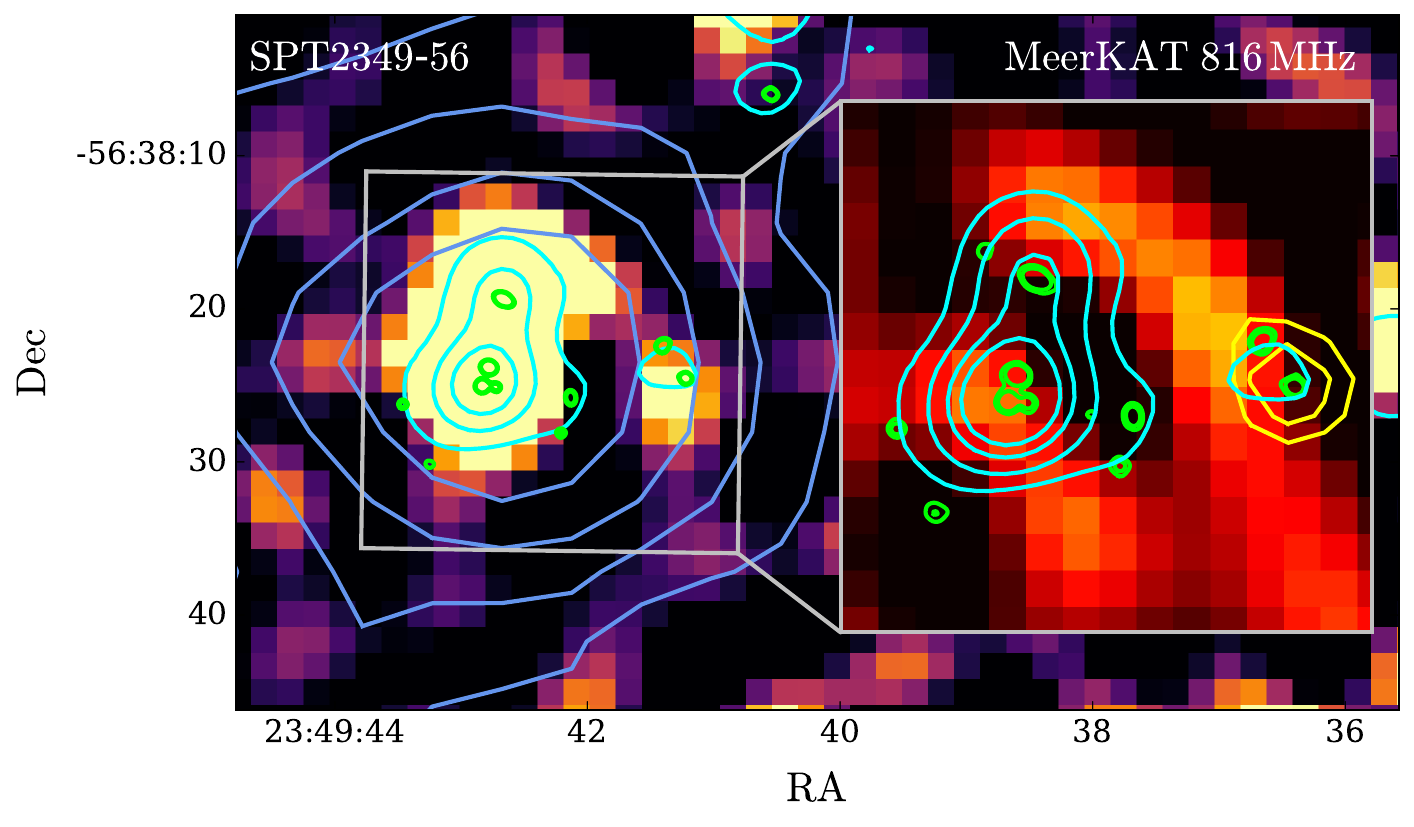}
    \caption{Zoomed in MeerKAT UHF maps with S-band contours (cyan) overlaid. The {\tt robust=-1.2} map is used for the north to maximize sensitivity to the well separated N1 and N2 sources. The {\tt robust=-2} is used for the core to cleanly separate source E from A and E. The protocluster region is highlighted by 870$\mu$m APEX-LABOCA contours (blue). {\bf INSETS:} Residual maps of the MeerKAT UHF images  are shown, after subtracting the five \spt\ sources: A, C, E, N1, N2. Protocluster sources detected by ALMA are identified with green contours in both the MeerKAT image and residual image. The MeerKAT maps are contoured over the residual maps for UHF (yellow) and S-band (cyan). For the core region (lower panel) we show in yellow contours only the residual `source C-subtracted' UHF map ({\tt robust=-1.2}) which reveals a 6.4$\sigma$ detection of source E.
    }
    \label{fig:radiosub}
\end{figure*}

The MeerKAT images used for flux measurements in Table~1 are shown in Figure~\ref{fig:radiosub}. Five source positions (A, C, E, N1, and N2) were modeled as unresolved sources in both S-band and UHF imaging, with peak positions allowed to vary. Subtracting these five sources yields residuals $<2\sigma$ as shown. 
The source positions do not vary by more than 1.5$''$ from the ALMA centroid in each case (i.e., well within the MeerKAT positional uncertainty, given the angular resolution and SNR of these detections). 

In the {\tt robust=-1.2} UHF 816\,MHz map, the central C, A, and E region is blended (Figure~\ref{fig:field}), however in the {\tt robust=-2} weighting, source E is cleanly separated, allowing a straightforward flux measurement (Table~1). 
An additional estimate of the UHF flux density for E was made using the more sensitive {\tt robust=-1.2} UHF map. Point source subtraction was performed on the UHF central region, removing the strong emission associated with ALMA source C. This map results in 
 identification of a strong 816\,MHz emission (6.3$\sigma$) centered on source E, shown in Figure~\ref{fig:field}. 
 There remains a possibility that the neighboring source D (S$_{850}$=4.5\,mJy) has radio emission roughly consistent with its SFR (expected to be 7$\mu$Jy at 2.4\,GHz; 15$\mu$Jy at 816\,MHz), and that source E is somewhat less radio-loud (7$\times$ compared to the 9.4$\times$ excess inferred if all the measured UHF radio flux is coming from E).

 For the source E, while the higher SNR {\tt robust=-1.2} UHF source extraction lies 0.4$''$ from the ALMA centroid, the S-band source extraction lies midway to the neighbouring source D (Figure~\ref{fig:radiosub}). Source D lies 3$''$ away from E, and the S-band source extraction centroid lies approximately 40\% towards D (1.2$''$ offset from E). Since D is expected to have $\sim$7$\mu$Jy S-band flux based on its far-IR luminosity, we instead extract peak S-band flux positions at the ALMA positions of sources E and D, finding 21$\pm5\mu$Jy at E and 12$\pm5\mu$Jy at D. The sum is consistent with the 35$\mu$Jy peak source subtraction above.
 
\subsection{Radio spectral indices}
We derive radio spectral indices directly for all five radio-detected \spt\ sources and list these  in Table~\ref{table:radio_full}. The data were fit according to a linear function in log-space using a Markov Chain Monte Carlo algorithm (MCMC) implemented by the \texttt{emcee} package~\citep{foreman2013}.  This MCMC package samples the posterior probability function, and is used to determine the error contours shown in Figure~\ref{fig:sed}, as well as the uncertainties on $\alpha$ in Table~\ref{table:radio_full}.  
 

\bibliographystyle{aasjournal}
\bibliography{SPT-PC}

\end{document}